\documentstyle[preprint,tighten,prl,aps,floats,epsfig]{revtex}
\begin{document}
\mbox{ }
\vspace{30mm}

\includegraphics{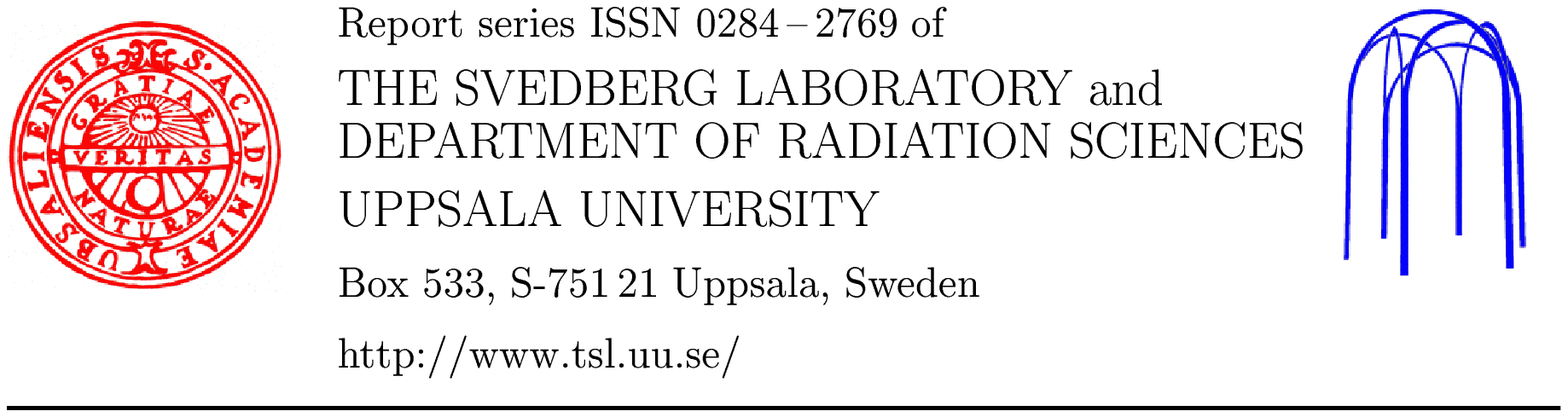}

\begin{flushright}
\begin{minipage}[t]{37mm}
{\bf TSL/ISV-99-0208 \\
May 1999} 
\end{minipage}
\end{flushright}

\vspace{5mm}

\begin{center}
{\LARGE\bf Has charge symmetry breaking been observed in the
$\bbox{dd \rightarrow\alpha\pi^0}$ reaction?}

\vspace{10mm}

{\Large Denis Dobrokhotov$^{\,a}$%
\footnote{ Electronic address: dvdobr@sci.lpi.msk.su},
G\"oran F\"aldt$^{\,b}$\footnote{ Electronic address: faldt@tsl.uu.se},
Anders G{\aa}rdestig$^{\,b}$\footnote{ Electronic address: grdstg@tsl.uu.se}, 
and Colin Wilkin$^{\,c}$\footnote{ Electronic address: cw@hep.ucl.ac.uk}}

\vspace{5mm}
{$^{a}$ Lebedev Physical Institute, 117924 Moscow, Russia}\\[2ex]
{$^{b}$ Division of Nuclear Physics, Uppsala University, Box 535,
S-751 21 Uppsala, Sweden}\\[2ex]
{$^{c}$ University College London, London WC1E 6BT, United Kingdom}
\end{center}

\begin{abstract}
Estimates are made of the $dd\to\alpha\gamma\gamma$ production cross
sections in a model where each neutron-proton pair in the beam and target
initiates an $np\to d\gamma$ reaction. This approach, which successfully
reproduces observables in two-pion production at intermediate energies,
suggests that direct two-photon production could provide a very significant
background to the measurement of the charge-symmetry-breaking (CSB) reaction
$dd\to \alpha\pi^0$. A non-vanishing CSB cross section has been reported
which might be confused with such two-photon production under the given
experimental conditions.\\
\end{abstract}

\clearpage
\setcounter{page}{1}

%\centerline{\today}
%\pacs{PACS: 11.30.Hv; 25.45.-z; 25.40.Lw; 25.10.+s}

\narrowtext

In quark language, the charge symmetry operator interchanges the $d$ and $u$ 
quarks and this almost leaves the system invariant because of the small mass 
differences between the current quarks~\cite{CSB}. These mass terms would, for 
example, mix different isospin states such as the $\pi^0$ and $\eta$ mesons. The
most convincing proof of charge symmetry breaking (CSB) in nuclear reactions 
would be the observation of a non-vanishing rate for the 
$dd\to \alpha\pi^0$ reaction~\cite{hen69}, because this would be proportional 
to the square of a CSB amplitude with no contribution from interference terms.

In a series of steadily more refined experiments~\cite{LG0,LG1,LG2}, a Saturne 
group first deduced upper bounds on the c.m.\ differential cross section, 
most notably $d\sigma/d\Omega \leq 0.8$~pb/sr at a beam energy of 
$T_d=800$~MeV~\cite{LG1} and a production angle of
$\theta_{\alpha}^{\text{lab}}=12^{\circ}$, corresponding to the peak of the 
Jacobian transformation from the c.m.\ to laboratory systems. The final
experiment was carried out at the higher energy of $T_d=1100$~MeV, also at 
$\theta_{\alpha}^{\text{lab}}=12^{\circ}$ 
($\theta_{\alpha}^{\text{cm}}=73^{\circ}$). This is no longer the Jacobian peak,
but was chosen to fit best the photon and $\alpha$-particle acceptances. The 
beam energy was taken close to the threshold for $\eta$ production in the 
analogous $dd\to \alpha\eta$ reaction (1121~MeV), with the hope that some 
$\eta$'s produced virtually might mix, through a CSB interaction, and emerge as 
$\pi^0$'s. A pion signal was claimed~\cite{LG2} with a c.m.\ cross section of 
\begin{equation}
 \frac{d\sigma}{d\Omega}(dd\to \alpha\pi^0) = 
 (1.0\pm 0.20\pm 0.25)~\mbox{\rm pb/sr}\:,
\end{equation}
where the first error is statistical and the second systematic. Within a simple
model~\cite{CW}, a cross section of around 1~pb/sr is consistent with other 
determinations of the $\pi^0/\eta$ mixing angle, although systematic 
uncertainties are difficult to quantify. 

The validity of the CSB interpretation was questioned at the time by some of the
experimentalists involved~\cite{PF} and, to understand the problem, a
description of the experiment is necessary. Well-identified $\alpha$ particles
were detected in the SPESIV magnetic spectrometer~\cite{SPES4}. The
scintillator hodoscope momentum binning of $\Delta p/p = 0.2\%$~\cite{LG1} was
degraded to a FWHM of 2\% through the opening of the collimator to increase the
counting rate. A \v{C}erenkov photon detector of 32 lead-glass blocks
had a good
acceptance for the two photons from $\pi^0$ decays. Each detected $\gamma$
was viewed by up to three neighboring blocks and its energy evaluated with a
precision of $\approx 30\%$ and its direction to within $\approx 3^{\circ}$.
The photon energy and angular information could be correlated with the 
$\alpha$-particle momentum on an event-by-event basis so that, apart from a
possible $e/\gamma$ ambiguity, the three final-state four-momenta of
$\alpha\gamma\gamma$ events could be determined.

The experiment resulted in 230 candidates with a detected $\alpha\gamma\gamma$ 
topology and 565 where only one photon was seen (although another photon could 
have escaped detection). By applying a series of severe cuts in the off-line 
analysis, the authors of Ref.~\cite{LG2} were left with 15 
$dd\to\alpha\gamma\gamma$ events which were spread in effective mass over 
the range $95\leq m_{\gamma\gamma}\leq 175$~MeV/$c^2$. Monte Carlo simulations 
indicate that such a wide spread in $m_{\gamma\gamma}$ is consistent
with single $\pi^0$ production and decay measured with a 2\% momentum 
resolution~\cite{LG2}. These 15 events correspond to a cross section of 
$(1.1\pm 0.30\pm 0.20)$~pb/sr. A slightly smaller figure came from 
analyzing a selection of the single-photon events, where one photon was 
presumed lost.

The alternative interpretation~\cite{PF} is that the $15$ events belong to a
continuum of $\alpha\gamma\gamma$ or $\alpha\gamma\gamma\gamma$ reactions, that
have been artificially selected by the experimental cuts. The population of
events compatible with the $\alpha\pi^0$ hypothesis does not show any obvious
accumulation in the plot of the $\alpha$-particle momentum {\it versus} the
$\gamma\gamma$ opening angle. This suggests that $\alpha\pi^0$ production 
is not the dominant process and, if it exists, is not separated in the 
data from the multi-photon continuum.

We have recently made estimates of two-pion production in the
$dd\rightarrow\alpha\pi^0\pi^0$ reaction in a model where both 
neutron-proton pairs undergo independent $np\to d\pi^0$ reactions, as indicated
in Fig.~1~\cite{GFW1,GFW2}. The predicted $2\pi^0$ cross section is roughly
proportional to the square of that for single $\pi^0$ production times a form
factor representing the probability for the two final deuterons sticking to form
an $\alpha$ particle. This overlap is very favourable because\ the c.m.\ frames
in the $np$ and $dd$ systems largely coincide. After inserting phenomenological
$np\to d\pi^0$ amplitudes, and including also the charged pion contribution, the
model reproduces well the observed differential cross section for
$dd\rightarrow\alpha X$ as a function of the missing mass $m_X$ and
$\alpha$-particle angle~\cite{GFW2}, as well as the measured deuteron vector and
tensor analyzing powers~\cite{Wurz}. The absolute normalization is reproduced to
within a factor of 1.5 throughout the 900 to 1300~MeV range of beam energies.
The predicted $2\pi^0$ cross section is shown in Fig.~2 under the conditions
corresponding to the CSB experiment~\cite{LG2}. This spectrum displays a
sharp ABC structure close to the $2\pi$ threshold~\cite{ABC}, as well as a
broader peak near the maximum missing mass. These features, which are equally
prominent in data~\cite{Ban76}, arise in our model from the shape of the 
$np\to d\pi^0$ cross section which is forward/backward peaked. The two-pion
cross section is then large when the two pions emerge parallel (the ABC peak) or
antiparallel (the central bump). We wish to use the same model to
estimate the background to the CSB experiment, replacing the $\pi^0$'s in
Fig.~1 by photons.

The formalism follows very closely that developed for two-pion
production~\cite{GFW2}, to which the reader is referred for further details. The
effective mass distributions in the $\alpha\gamma\gamma$, or 
$\alpha\pi^0\pi^0$ channels are expressed in terms of the matrix element 
${\cal M}$ through
\widetext
\begin{equation}
 \frac{d^2\sigma}{d\Omega\,dm_X}(dd\to\alpha X) = 
 \frac{1}{64(2\pi)^5} 
 \frac{k_{\alpha}^{\text{cm}}k^{\ast}}{p_{d}^{\text{cm}}s}
 \frac{1}{9} \int d\Omega^{\ast}\sum_{\text{ext pol}} 
 |{\cal M}^{\ast}|^2,
\end{equation}
\narrowtext
\noindent
where $\sqrt{s}$ is the total c.m.\ energy, $k$ the relative $\gamma\gamma$ or
$\pi^0\pi^0$ momentum, and the sum is over external spin projections. Quantities
denoted by an asterisk ($\ast$) are evaluated in the $2\gamma$ or $2\pi$ rest
frame. 

In evaluating the amplitudes corresponding to Fig.~1, we neglect the
deuteron D-state and the influence of the Fermi motion on the spin 
couplings. The matrix element then factorizes into a kernel $\cal K$, 
that contains the spin couplings, and a form factor 
$\cal W$; ${\cal M}=-i(m_{\alpha}/v_{d})\,{\cal K\cal W}$, where $v_d$
is the deuteron speed. In this approximation the form of $\cal W$ is the 
same as that for two-pion production, which is successfully described by this
approach~\cite{GFW2}.

Rather than averaging the c.m.\ energies of the sub-processes over the Fermi
momenta, these are fixed by assuming that the two production reactions share
the total c.m.\ energy equally. For two-photon production this assumption means
that the photon laboratory energy in the inverse $\gamma d\to np$ reaction
is given by $E_{\gamma}^{\text{lab}}=T_d/4$.

The input $np\to d\gamma/\pi^0$ vertices are parametrized in terms of
experimentally-determined partial-wave amplitudes, with those for pion
production being discussed in our earlier work~\cite{GFW2}. The photon
amplitudes were obtained from the phenomenological multipoles of
Arenh\"ovel~\cite{Aren} using
\widetext
\begin{equation}
 AL(^{2s+1}l_j) = \sqrt{\frac{2l+1}{4\pi(2j+1)}}\sum_{m_s,m_d}
 \langle 1m_d\,Lm_s-m_d|jm_s\rangle \langle l0\,sm_s|jm_s \rangle
 \langle sm_s|A^{(L)}|m_d \rangle,
\label{eq:amprel}
\end{equation}
\narrowtext
\noindent
where $\langle sm_s|A^{(L)}|m_d \rangle$ is a spin-projected multipole. In this
notation ($L,j$) are the total angular momenta of the photon and the whole
process, ($s,l$) the spin and orbital angular momentum of the $np$ system, and
$A$ denotes either electric ($E$) or magnetic ($M$). At
$E_{\gamma}^{\text{lab}}= T_d/4 = 275$~MeV the $np\to\gamma d$ cross section is
dominated by $M1(^1\!D_2)$ and $E1(^3\!F_2)$ partial waves~\cite{Aren2}, that
have angular distributions $(3\sin^2\theta+2)$ and $(\sin^2\theta+1)$
respectively and which are both maximal at $90^{\circ}$. Together these
amplitudes reproduce most of the observed $(a\sin^2\theta+1)$ behavior, with
$a>1$. This variation is to be contrasted with the $(3\cos^2\theta+1)$ 
dependence, typical for $np\to\pi^0 d$ at these energies, which 
is sharply peaked towards $\theta=0^{\circ}$ and $180^{\circ}$.

In the spin-amplitude formalism~\cite{GFW2}, the spin structure of the 
dominant partial waves is given by
\widetext
\begin{equation}
 M1(^1\!D_2) \leftrightarrow -\frac{i\sqrt{3}}{2} \left\{
 (\bbox{\hat{p}}\cdot\bbox{\epsilon}_{d}^{\dagger})
 \bbox{\hat{p}}\cdot(\bbox{\hat{k}\times\epsilon}_{\gamma}^{\dagger})
 -\frac{1}{3}
 \bbox{\epsilon}_{d}^{\dagger}\cdot
 (\bbox{\hat{k}\times\epsilon}_{\gamma}^{\dagger})
 \right\},
\end{equation}
\begin{equation}
 E1(^3\!F_2) \leftrightarrow -\frac{\sqrt{5}}{2} \left\{
 \bbox{\sigma}\cdot\bbox{\hat{p}} \left[ 
 (\bbox{\hat{p}}\cdot\bbox{\epsilon}_{d}^{\dagger})
 (\bbox{\hat{p}}\cdot\bbox{\epsilon}_{\gamma}^{\dagger})
 -\frac{1}{5}\bbox{\epsilon}_{d}^{\dagger}
 \cdot\bbox{\epsilon}_{\gamma}^{\dagger} \right] \right.
 \left. -\frac{1}{5} \left[
 (\bbox{\sigma}\cdot\bbox{\epsilon}_{d}^{\dagger})
 (\bbox{\hat{p}}\cdot\bbox{\epsilon}_{\gamma}^{\dagger})+
 (\bbox{\sigma}\cdot\bbox{\epsilon}_{\gamma}^{\dagger})
 (\bbox{\hat{p}}\cdot\bbox{\epsilon}_{d}^{\dagger}) 
 \right] \right\},
\end{equation}
\narrowtext
\noindent
where ($\bbox{p},\bbox{k}$) are the proton and photon momenta, 
($\bbox{\epsilon}_{d}^{\dagger},\bbox{\epsilon}_{\gamma}^{\dagger}$) the 
deuteron and photon polarisation vectors, and $\bbox{\sigma}$ the Pauli spin
matrices. These expressions are to be multiplied by the corresponding complex
amplitude determined using the spin structure of Eq.~(\ref{eq:amprel}).

In a first approach, only the two $E1$ and $M1$ photodisintegration amplitudes
were retained. The predictions for the $\alpha\gamma\gamma$ and
$\alpha\pi^0\pi^0$ channels are given in Fig.~2. The magnitude of the
two-photon cross section is sensitive to both the relative strength and phase of
the two photon-producing amplitudes, with values ranging from 0.4 to 1.3 times
the predicted curves being possible with reasonable variation of these two
parameters.

The solid curves in Fig.~2 represent the result of a calculation using all the
$np\to d\gamma$ amplitudes with $L\leq 2$, as determined by
Arenh\"ovel~\cite{Aren2}. These amplitudes predict a $\gamma d\to np$ total
cross section about a factor of 1.2 larger than that suggested by the available
data~\cite{Whis} and so the input was reduced by this amount. The similarity of
the two calculations illustrated in Fig.~2 indicates that the small $np\to
d\gamma$ amplitudes are not crucial in our estimates. The complete lack of
structure in the $2\gamma$ effective mass distributions is in stark contrast to
the ABC peaks in the $2\pi$ spectrum. This is a direct consequence of the very
different angular dependences of the subprocesses and, in particular, the
tendency for the photons to emerge at large c.m.\ angles. There is, however, a
strong angular dependence in the $\gamma\gamma$ form factor. For 
$\alpha$-particle momenta $k_{\alpha}^{\text{cm}} > 420$~MeV/$c$, corresponding
to  $m_X < 300$~MeV/$c^2$, photon emission parallel to the $\alpha$ particle
is expected to be orders of magnitude smaller than perpendicular emission.

The model does indeed predict the production of a significant $2\gamma$
continuum. Its gross contribution in the $\pi^0$ region may be estimated by
integrating the missing mass distribution in Fig.~2 over the experimental
acceptance interval $95<m_X<175~\text{MeV}/c^2$~\cite{LG2}. This leads
to a c.m.\ cross section of
$d\sigma(\alpha\gamma\gamma)/d\Omega \approx 8.8$~pb/sr at
$\theta_{\alpha}^{\text{cm}}=73^{\circ}$. Though this is larger than the 
$\pi^0$-signal reported~\cite{LG2}, the reduction resulting from the
experimental cuts imposed is hard to quantify. If the $2\gamma$ distribution 
were uniform, simple cuts would reduce the signal by a factor
of three or more~\cite{PF}, though this might be modified by any strong photon 
angular distribution. The naive initial flux damping factor introduced in
Ref.~\cite{GFW2}, could also diminish the cross section by $\approx 10-15\%$. 
After taking such reductions into account, the similarity between our estimate
and the claimed $\pi^0$ signal~\cite{LG2} gives cause for concern, especially
since there is a theoretical uncertainty of at least a factor of two, due in 
part to errors in the photoproduction input. Unless the influence of the 
predicted $\gamma\gamma$ continuum can be reduced, the significance of the CSB 
measurement must be questioned.

Estimates at $T_d=800$~MeV give a similar value for the double-differential
cross section of
$d^2\sigma(\alpha\gamma\gamma)/d\Omega\,dm_{\gamma\gamma} \approx 
0.11$~pb/(sr MeV/$c^2$). The early experiment~\cite{LG1} quoted only an upper 
bound for $\pi^0$ production because of an unidentified but significant
background that varied smoothly with angle. Taking into account the effect of
the cut imposed on the basis of the \v{C}erenkov information, the
background cross section at $\theta_{\text{lab}}=12^{\circ}$ was 
$(5\pm 2)$~pb/sr. Integrating over an experimental acceptance of 
$\approx 40$~MeV/$c^2$ yields a $2\gamma$ estimate of almost exactly 
this figure, suggesting that the background is indeed due to two-photon
production as discussed here.

Another possible background to the CSB experiment might arise from a
$\pi^0\,\gamma$ final state with one very soft (and undetected) photon. One
could try to estimate this cross section in a similar model to that
of Fig.~1, replacing just one of the pions by a soft photon. However, in this
kinematic limit the momentum sharing is destroyed and the form factor
${\cal W}$ becomes very small and model dependent. Within our approach 
the $\pi^0\,\gamma$ background is likely to be far less serious than the
$2\gamma$ one studied here.

Our calculations suggest that the evidence for charge symmetry violation in the
$dd \rightarrow\alpha\pi^0$ reaction~\cite{LG2} must be treated with great
caution. We have shown that direct two-photon production is important
for both this and the earlier experiment~\cite{LG1}. Experiments are
needed with a better $\pi^0$ mass resolution and, though the signal might 
be weaker, this is most easily achieved near the $\alpha\pi^0$ threshold 
($T_d = 226$~MeV). At say 10~MeV above threshold in the c.m., the 
dominant photodisintegration amplitudes are $E1(^3\!P_1)$ and 
$E1(^3\!P_2)$~\cite{Aren2} and we predict an integrated
cross section of $d\sigma(\alpha\gamma\gamma)/dm_{\gamma\gamma} 
= 3.4~\mbox{pb/(MeV/$c$}^2)$. At the IUCF storage ring the experiment
could be carried out with tensor polarised deuterons~\cite{bacher}.
Pion production at threshold has an analyzing power of
$t_{20}=1/\sqrt{2}$~\cite{Wurz}, whereas our prediction for two-photon
production gives $t_{20}$ consistent with zero. Alternatively,  
a new experiment might take advantage of the
distribution in the angle $\theta_{\gamma\gamma}$ of the two photons in their 
rest frame, which must be isotropic for true $\pi^0$ production. Because of 
the preferential photon emission at $90^{\circ}$ in the $np\to d\gamma$ 
reaction at low energies, our model suggests that, near threshold and in the
forward $\alpha$-particle direction,
$d\sigma/d\Omega_{\gamma\gamma} \propto
(1+b\sin^2\theta_{\gamma\gamma})^2$, where $b \approx 2$. This could
be investigated with the WASA $4\pi$ $\gamma$-detector at the CELSIUS
ring~\cite{WASA}.

One of the authors (C.W.) gratefully acknowledges many detailed discussions over
15 years with L.~Goldzahl regarding the CSB experiments described here. The
crucial importance of direct $2\gamma$ production, possibly from the decay of
two $\Delta$'s, was pointed out to us by P.~Fleury and F.~Plouin. H.~Arenh\"ovel
has been very helpful in providing the necessary photodisintegration 
amplitudes. This work has benefitted from the continued financial support
of the Swedish Royal Academy and the Swedish Research Council. Two of the 
authors (D.D. \&\ C.W.) wish to thank the The Svedberg Laboratory for its 
generous hospitality.

%
% Figure captions
%
\mbox{}
\vspace{30mm}
\begin{figure}[p]
\includegraphics{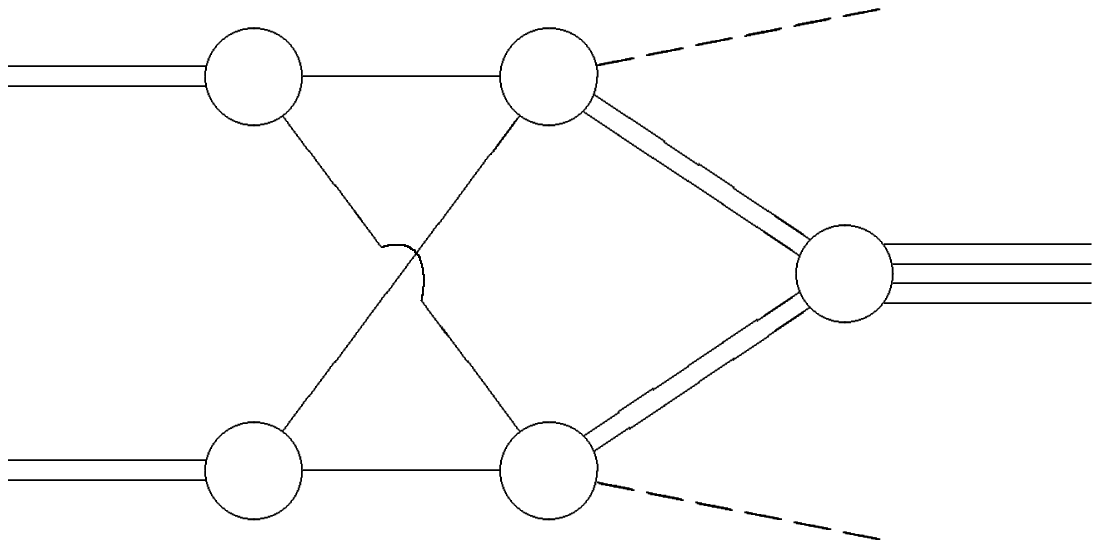}
\caption{Feynman diagram for the $dd\to\alpha\gamma\gamma$ and 
$dd\to\alpha\pi^0\pi^0$ reactions, where the dashed lines denote either
pions or photons.}
\end{figure}
\clearpage
\begin{figure}[p]
\includegraphics{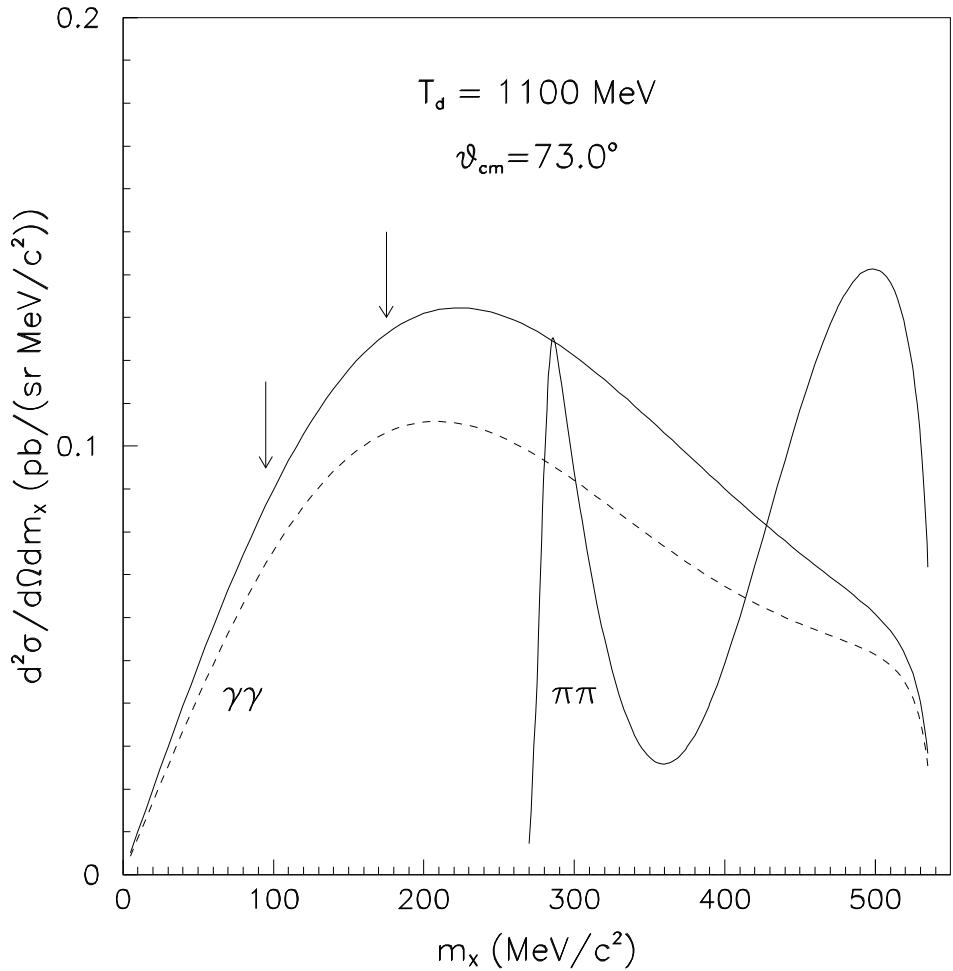}
\vspace{100mm}
\caption{Predicted missing mass distributions at $T_d=1100$~MeV and 
$\theta_{\alpha}^{\text{cm}}=73^{\circ}$ for the 
$dd\to\alpha\gamma\gamma$ and $dd\to\alpha\pi^0\pi^0$ reactions, the 
latter being reduced by a factor of $10^4$. Solid lines in the photon case 
correspond to the full calculation, incorporating all $L\leq 2$ amplitudes, 
while the dashed lines are derived purely from the $M1(^1\!D_2)$ and 
$E1(^3\!F_2)$ amplitudes. The limits of the experimental acceptance (95 to 
175~MeV/$c^2$) are indicated. }
\end{figure}
\end{document}